\documentclass[%
 reprint,
 superscriptaddress,
 amsmath,amssymb,
 aps,
prb,
]{revtex4-1}

\usepackage{comment, color}
\usepackage{braket}

\usepackage{graphicx}
\usepackage{dcolumn}
\usepackage{bm}

\begin{document}

\title{Topological exceptional surfaces in non-Hermitian systems with parity-time and parity-particle-hole symmetries}

\author{Ryo Okugawa}
 \affiliation{%
 	WPI-Advanced Institute for Materials Research (WPI-AIMR), Tohoku University, 2-1-1, Katahira, Sendai 980-8577, Japan
 }%
\author{Takehito Yokoyama}%
 \affiliation{%
 	Department of Physics, Tokyo Institute of Technology, 2-12-1 Ookayama, Meguro-ku, Tokyo 152-8551, Japan 
}%

\date{\today}

\begin{abstract}
We study a topological band degeneracy in non-Hermitian systems with parity-time ($PT$) and parity-particle-hole ($CP$) symmetries. 
In $d$-dimensional non-Hermitian systems,
it is shown that $(d-1)$-dimensional exceptional surfaces can appear from band touching thanks to $PT$ or $CP$ symmetry.
We investigate the topological stability and zero-gap quasiparticles for the exceptional surfaces due to the band degeneracy.
We also demonstrate the band degeneracy by using lattice models of a topological semimetal and a superconductor.
\end{abstract}

\maketitle

\section{Introduction}
Since the discovery of topological insulators \cite{Hasan10, Qi11},
many topological gapped phases have been suggested from systematic topological classification of quantum matters \cite{Chiu16}.
This idea of the classification is generalized to gapless phases \cite{Matsuura13, Zhao13, Zhao14}.
Topologically gapped and gapless phases are related
because some topological gapless phases necessarily intervene between trivial and topological gapped phases
\cite{Murakami07, Murakami08, Burkov11, Okugawa14, Liu14, Alexandradinata14, Yang14, Kim16, Ahn17, Ahn18, Meng12, Okugawa18}.
Furthermore, crystal symmetries play an important role in topological phases.
For example,
parity ($P$) symmetry enables different topological phases \cite{Hughes11, Turner12, Kim15, Yu15, Fang15, Zhao16, Bzdusek17, Khalaf18}.

The concept of the topological phases has been recently extended to non-Hermitian systems \cite{Martinez18}.
While one-dimensional non-Hermitian topological gapped phases have been mainly studied
\cite{Rudner09, Zhu14, Zeuner15, Malzard15, Yuce15, Lee16, Weimann17, Martinez18B, Yuce18, Yao18L, Lieu18B, Yin18},
more recent works have investigated nontrivial phases in two- and three-dimensional non-Hermitian systems
\cite{Esaki11, Sato12, Shen18, Gong18, Yao18, Kawabata18, Kawabata18a, Xiong18, Chen18, Philip18}.
Novel topological phases are allowed in high-dimensional systems \cite{Gong18, Kawabata18}
because time-reversal ($T$) and particle-hole ($C$) symmetries are unified by non-Hermiticity
\cite{Sato12, Peng16, Gong18, Kawabata18}.
As with Hermitian systems, particle-hole symmetry can be seen in various non-Hermitian topological systems
\cite{Malzard15, Wang15, Yuce16, Klett17, Menke17, Yuce18, Lieu18B, Kawabata18, KawabataB, Lieu18}.
Non-Hermiticity also provides relevant topological phenomena
such as topological insulator lasers \cite{Harari18, Bandres18} and anomalous edge states \cite{Lee16, Xiong18, Martinez18B, Kawabata18a}.

As non-Hermitian topological gapless structures,
exceptional points emerge from band degeneracy in two-dimensional systems
\cite{Berry04, Heiss12, Dembowski01, Mailybaev05, Leykam17, Kozii17, Zhou18, Papaj18, Zhao18, Yoshida18, Shen18B}.
The exceptional points appear 
between normal and Chern insulator phases in non-Hermitian systems \cite{Shen18, Yao18, Kawabata18a}.
In three-dimensional systems,
exceptional lines are realizable in the momentum space \cite{Xu17, Cerjan18, Zyuzin18, Carlstrom18, Yang18, Wang18, Cerjan18a}.
In addition to topological phases, such non-Hermitian band touchings give zero-gap quasiparticles
\cite{Kozii17, Zhou18, Papaj18, Zhao18, Yoshida18, Shen18B, Zyuzin18, Carlstrom18, Yang18}.
For instance, the zero-gap states form lines called bulk Fermi arcs in the two-dimensional momentum space
\cite{Kozii17, Zhou18, Papaj18, Zhao18, Yoshida18, Shen18B}.
Moreover, when non-Hermitian systems have parity-time ($PT$) symmetry,
exceptional points and lines can emerge in the one-dimensional \cite{Malzard15, Ding15, Lieu18B}
and the two-dimensional bands \cite{Szameit11, Ramezani12, Zhen15, Cerjan16}, respectively.
Hence, it is beneficial to reveal a relationship between the band degeneracy and symmetry
through non-Hermitian topology in general dimension.

In this paper, we study band degeneracies such as exceptional lines and exceptional surfaces in non-Hermitian systems
when parity-time ($PT$) or parity-particle-hole ($CP$) symmetry is present.
In particular, the cases of $(PT)^2=1$ and $(CP)^2=1$ are investigated.
We elucidate generic conditions to realize exceptional surfaces, and characterize them by using $\mathbb{Z}_2$ topological invariants.
It is shown that zero-gap quasiparticle states accompany the topological exceptional surfaces. 
To confirm our theory, we investigate non-Hermitian lattice models of a topological nodal-line semimetal and an even-parity superconductor.

This paper is organized as follows.
In Sec.~\ref{BD}, we construct effective models to describe exceptional surfaces in non-Hermitian systems.
In Sec.~\ref{Z2}, we discuss general topological properties of the exceptional surfaces.
We study lattice models to show the exceptional surfaces in Sec.~\ref{lattice}.
Our conclusion is given in Sec.~\ref{conc}.

\section{Band degeneracy in non-Hermitian systems}\label{BD}
In this section, we study general conditions to realize exceptional points, lines and surfaces.
We investigate an accidental band degeneracy between two states in non-Hermitian systems.
Therefore, we consider two-band non-Hermitian effective models.

\subsection{General two-band effective Hamiltonian}
Firstly, we introduce the topological theory of band degeneracy in non-Hermitian systems without symmetry.
When two states are nondegenerate, the energy eigenvalues can be described by a $2\times 2$ matrix which represents the subspace of the two states \cite{Berry04, Heiss12, Shen18}.
We apply the theory to non-Hermitian bands with $PT$ and $CP$ symmetries in the next section.
The generic effective Hamiltonian at wavevector $\bm{k}$ is 
\begin{align}
	H(\bm{k})=(a_0+ia_1)\sigma _0+\bm{b}_0\cdot \bm{\sigma}+i\bm{b}_1\cdot \bm{\sigma}, \label{2deff}
	\end{align}
where $\sigma _{0}$ and $\bm{\sigma}=(\sigma _x,\sigma _y,\sigma _z)$ are the identity matrix and Pauli matrices, respectively.
Here, $a_{0,1}$ and $\bm{b}_{i=0,1}=(b_{i,x},b_{i,y},b_{i,z})$ are real.
The energy eigenvalues are 
\begin{align}
	E_{\pm}(\bm{k})=a_0+ia_1\pm \sqrt{\bm{b}_0\cdot \bm{b}_0-\bm{b}_1\cdot \bm{b}_1+2i\bm{b}_0\cdot \bm{b}_1}.
	\end{align}
If the two states become degenerate, i.e., $E_+=E_-$, the bands satisfy 
\begin{align}
	\bm{b}_0\cdot \bm{b}_0-\bm{b}_1\cdot \bm{b}_1=0, \hspace{5mm}
	 \bm{b}_0\cdot \bm{b}_1=0. \label{cond2}
	\end{align}
Therefore, we have the two conditions for the band degeneracy.
To satisfy these conditions, the system needs (at least) two tunable parameters.
In general, structures of band degeneracy are determined by the difference
between the number of parameters and that of conditions for the band degeneracy. 
In $d$-dimensional non-Hermitian systems, since we have $d$ components of the wavevector,
band degeneracy with $(d-2)$-dimensional surface is realizable if $d\geq 2$ [Figs.~\ref{ex}(a) and \ref{ex}(b)].
The band degeneracy can be characterized by a half-integer topological invariant \cite{Shen18, Zhou18, Wang18}.
Note that the conditions in Eq.~(\ref{cond2}) are unchanged by only unitary crystal symmetry $UHU^{-1}=H$ with a unitary matrix $U$
because this symmetry equally acts on $\bm{b}_0\cdot \bm{\sigma}$ and $\bm{b}_1\cdot \bm{\sigma}$ in Eq.~(\ref{2deff}).
Thus, we consider anti-unitary symmetries $PT$ and $CP$.

\subsection{Exceptional surfaces in $PT$- and $CP$-symmetric non-Hermitian systems}
\begin{figure}[t]
	\includegraphics[width=6.5cm]{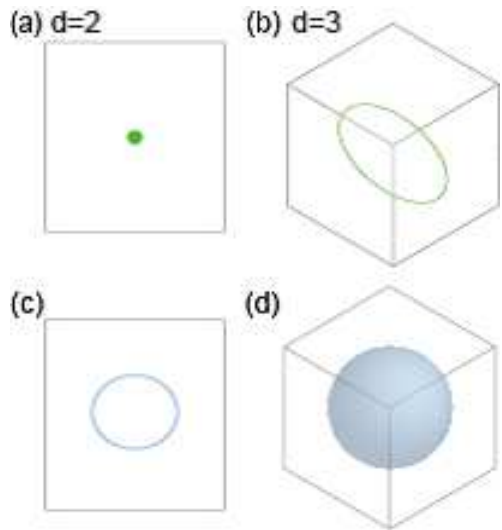}
	\caption{\label{ex} Schematic drawings of band degeneracies in the non-Hermitian systems. (a) An exceptional point and (b) an exceptional line in the two- and the three-dimensional systems without symmetries. (c) An exceptional line and (d) an exceptional surface in the two- and the three-dimensional systems with $PT$ or $CP$ symmetry.
	}
\end{figure}

We construct effective models to clarify conditions of band degeneracies in $PT$- and $CP$-symmetric non-Hermitian systems.
We study both cases of $(PT)^2=1$ and $(CP)^2=1$.
Firstly, we consider $PT$ symmetry which imposes $(PT)H(\bm{k})(PT)^{-1}=H(\bm{k})$ on the Hamiltonian in Eq.~(\ref{2deff}).
Because this symmetry is antiunitary,
it can be generally represented as $PT=UK$, where $U$ is a unitary matrix and $K$ is complex conjugation.
From $(PT)^2=1$, we can set $U=\sigma _0$ by choosing a proper basis \cite{Sato12}.
Then, the effective Hamiltonian is
\begin{align}
	H(\bm{k})=a_0\sigma _0+b_{0,x}\sigma _x+b_{0,z}\sigma _z+ib_{1,y}\sigma _y. \label{PTH}
	\end{align}
The energy eigenvalues are 
\begin{align}
	E_{\pm}(\bm{k})=a_0\pm \sqrt{\bm{b}_0\cdot \bm{b}_0-\bm{b}_1\cdot \bm{b}_1}. \label{PTenergy}
\end{align}
Hence, $PT$ symmetry automatically leads to $\bm{b}_0\cdot \bm{b}_1=0$ in Eq.~(\ref{cond2}).
In other words, the band degeneracy happens if $\bm{b}_0\cdot \bm{b}_0-\bm{b}_1\cdot \bm{b}_1=0$ is satisfied in the momentum space.
The number of conditions for the band degeneracy is reduced to one.
As a result, $(d-1)$-dimensional exceptional surfaces can emerge in the $d$-dimensional non-Hermitian systems when $d\geq 1$.
Thus, as shown in Figs.~\ref{ex}(c) and (d), we can obtain exceptional lines and surfaces in two- and three-dimensional non-Hermitian systems with $PT$ symmetry, respectively.

In the presence of the $PT$ symmetry, energy eigenvalues can be real \cite{Bender98, Bender07, Konotop16}.
Then, we can see that the number of real eigenvalues changes on the exceptional surface.
From Eq.~(\ref{PTenergy}), 
we have two real energy eigenvalues in the momentum space for $\bm{b}_0\cdot \bm{b}_0>\bm{b}_1\cdot \bm{b}_1$.
Meanwhile, the eigenvalues become complex in the region $\bm{b}_0\cdot \bm{b}_0<\bm{b}_1\cdot \bm{b}_1$.
The change of the number of real eigenvalues is significant for the topological characterization of the exceptional surfaces,
as described in Sec.~\ref{Z2}.

Secondly, let us consider $CP$ symmetry described by $(CP)H(\bm{k})(CP)^{-1}=-H(\bm{k})$.
$CP$ symmetry is also anti-unitary.
When $CP=K$ is chosen in a proper basis, we obtain
\begin{align}
	H(\bm{k})=ia_1\sigma _0 +b_{0,y}\sigma _y+ib_{1,x}\sigma _x+ib_{1,z}\sigma _{z}. \label{CPH}
	\end{align}
Therefore, $CP$ symmetry also gives $\bm{b}_0\cdot \bm{b}_1=0$.
Hence, we can also obtain $(d-1)$-dimensional exceptional surfaces in the $d$-dimensional $CP$-symmetric systems.
Moreover, it is found that the number of purely imaginary eigenvalues changes on the exceptional surface in the $CP$-symmetric system.

\section{Topological properties of exceptional surfaces} \label{Z2}
Here, we characterize exceptional surfaces topologically when non-Hermitian systems have $PT$ or $CP$ symmetry.
We also reveal the topological properties associated with the exceptional surfaces.
For simplicity, we assume that the exceptional surfaces exist at zero energy without loss of generality
because the origin of the energy can be shifted.
In this section, we refer to $(d-1)$-dimensional exceptional surfaces just as exceptional surfaces.

\subsection{$\mathbb{Z}_2$ topological characterization for exceptional surfaces}
We begin with topological characterization of exceptional surfaces in $PT$ symmetric non-Hermitian systems.
Before going into the details, we introduce a topological invariant for the zero-dimensional non-Hermitian systems with $PT$ symmetry, according to Ref.~\onlinecite{Gong18}.
The $PT$-symmetric Hamiltonians $H_{PT}$ are classified by a $\mathbb{Z}_2$ topological invariant given by \cite{Gong18}
\begin{align}
	s=\mathrm{sgn} \det (H_{PT})=(-1)^{n_{R_-}}, \label{PTinv}
	\end{align}
where $n_{R_-}$ is the number of the real negative energy eigenvalues.
The topological invariant $s$ is defined when $H_{PT}$ does not have zero-energy eigenvalues.
If $s=+1(-1)$, the system has an even (odd) number of energy eigenvalues on the negative energy axis \cite{Gong18}.
Therefore, when the system has zero-energy eigenvalues, it can be understood as a topological phase transition.

When a non-Hermitian $PT$-symmetric system has a band degeneracy,
the momentum space is divided by the exceptional surfaces, as illustrated in Figs.~\ref{ex}(c) and (d).
To characterize the exceptional surfaces,
we regard each point $\bm{k}$ in the momentum space as a zero-dimensional system with $PT$ symmetry.
Because the Hamiltonian satisfies $(PT)H(\bm{k})(PT)^{-1}=H(\bm{k})$ for all $\bm{k}$,
we can define $s(\bm{k})$ on each point by using Eq.~(\ref{PTinv}).
By assumption, we can interpret the exceptional surface as a topological phase transition in the momentum space by changing $\bm{k}$.
Thus, the exceptional surfaces can be characterized by the $\mathbb{Z}_2$ topological invariant leading to topological protection \cite{comment}.
This argument is consistent with the change of $n_{R_-}(\bm{k})$ on the exceptional surface, as discussed in Sec.~\ref{BD} B.

Next, we consider $CP$-symmetric non-Hermitian systems. 
Similarly, zero-dimensional $CP$-symmetric systems are classified by a $\mathbb{Z}_2$ topological invariant \cite{Gong18}.
Thus, the $\mathbb{Z}_2$ topological invariant can characterize exceptional surfaces in the $CP$-symmetric systems.
For the zero-dimensional $CP$-symmetric Hamiltonian $H_{CP}$, the $\mathbb{Z}_2$ topological invariant is \cite{Gong18}
\begin{align}
	s'=\mathrm{sgn}\det(iH_{CP})=(-1)^{n_{I_+}},
	\end{align}
where $n_{I_+}$ is the number of the energy eigenvalues on the positive imaginary axis.
We define the topological invariant $s'$ when $H_{CP}$ does not have zero-energy eigenvalues.
Eventually, we can also characterize the exceptional surfaces in a similar way to $PT$-symmetric systems
if we define $s'(\bm{k})$ for each point $\bm{k}$ in the momentum space.

In both the $PT$- and the $CP$-symmetric systems,
the exceptional surfaces can be characterized by the $\mathbb{Z}_2$ topological invariants.
The same $\mathbb{Z}_2$ classifications stem from 
topological unification of time-reversal and particle-hole symmetries in the non-Hermitian gapped systems \cite{Gong18, Kawabata18}.
By the transformation $H \rightarrow iH$, 
the two symmetries can be actually treated as one symmetry called $K$ symmetry in non-Hermitian systems \cite{Bernard02, Ulrika08, Sato12}.
As a result, the topological unification can be found in the characterization of the exceptional surfaces.

\subsection{Topological stability and drumhead bulk states}
In order to see the topological stability,
we consider a simple exceptional line in the two-dimensional system.
The results in this section can be generalized to exceptional surfaces.
Suppose that the exceptional line is described by 
\begin{align}
	H(\bm{k})=k_x\sigma _x+k_y\sigma _y+i\delta \sigma _z, \label{2ddsm}
\end{align}
where $\delta$ is a real constant.
This model can be regarded as a two-dimensional Dirac cone with the non-Hermitian term $i\delta \sigma _z$,
and realized in a honeycomb photonic crystal with gain and loss \cite{Szameit11, Ramezani12, Zhen15}.
The Hamiltonian has $PT$-symmetry represented by $PT=\sigma_x K$.
(If we perform the transformation $H \rightarrow iH$, the model becomes $CP$-symmetric.)
The eigenvalues are
\begin{align}
	E_{\pm}(\bm{k})=\pm \sqrt{k_x^2+k_y^2-\delta ^2}.
\end{align}
Therefore, the exceptional line appears on $k_x^2+k_y^2=\delta ^2$.
Then, the topological invariant for the exceptional line is 
\begin{align}
	s(\bm{k})=
	\begin{cases}
		+1  &(k_x^2+k_y^2 < \delta ^2) \\
		-1 &(k_x^2+k_y^2 > \delta ^2)
	\end{cases}.
\end{align}
Thanks to the topological protection,
the exceptional line is stable as long as the Hamiltonian preserves the $PT$ symmetry.

Indeed, the exceptional line disappears by perturbations breaking the $PT$ symmetry of the Hamiltonian.
If we add the $\delta '\sigma _z$ term with a real constant $\delta '$ to Eq.~(\ref{2ddsm}),
the exceptional line vanishes because the eigenvalues become $E_{\pm}=\pm \sqrt{k_x^2+k_y^2+\delta '^2-\delta^2+2i\delta '\delta}$.
On the other hand, if the $i\kappa _x\sigma _x+i\kappa _y\sigma _y$ term is added,
the exceptional line can become exceptional points \cite{Shen18, Kozii17}.
When $\kappa _x$ and $\kappa_y$ are real, the exceptional points emerge at
$(k_x, k_y)=\pm (-\kappa _y \sqrt{\frac{\delta ^2+\kappa _x^2+\kappa _y^2}{\kappa _x^2+\kappa _y^2}},\kappa _x \sqrt{\frac{\delta ^2+\kappa _x^2+\kappa _y^2}{\kappa _x^2+\kappa _y^2}})$.

From the topological property,
the quasiparticle band gap becomes zero on a region bordered by the exceptional line.
The quasiparticle band gap is given by $\Delta _{q} (\bm{k})=\mathrm{Re}E_{+}-\mathrm{Re}E_{-}$
in non-Hermitian systems \cite{Kozii17}.
In this model, we can write the band gap as
\begin{align}
	\Delta _{q}(\bm{k})=
	\begin{cases}
		0  &(k_x^2+k_y^2 \leq \delta ^2) \\
		2\sqrt{k_x^2+k_y^2-\delta ^2} &(k_x^2+k_y^2 > \delta ^2)
	\end{cases}.
	\end{align}
Because $\Delta _{q}(\bm{k})=0$ inside the exceptional line, the drumheadlike gapless quasiparticles appear in the bulk.
Although the drumhead bulk states are analogous to drumhead surface states of nodal-line semimetals \cite{Kim15, Yu15},
these zero-gap bulk states can be found since the eigenvalues are purely imaginary inside the exceptional lines.

In general, we can see that a quasiparticle band gap can vanish in the regions surrounded by exceptional surfaces as follows.
From the energy eigenvalue of the effective model in Eq.~(\ref{PTenergy}),
we obtain
\begin{align}
	\Delta _{q} (\bm{k})=
	(1+\mathrm{sgn}(\bm{b}_0\cdot \bm{b}_0-\bm{b}_1\cdot \bm{b}_1))\sqrt{\bm{b}_0\cdot \bm{b}_0-\bm{b}_1\cdot \bm{b}_1}.
\end{align}
Therefore, the zero-gap states can appear when $\mathrm{sgn}(\bm{b}_0\cdot \bm{b}_0-\bm{b}_1\cdot \bm{b}_1)$ changes.
Since the change of $\mathrm{sgn}(\bm{b}_0\cdot \bm{b}_0-\bm{b}_1\cdot \bm{b}_1)$ corresponds to that of $s(\bm{k})$,
the zero-gap quasiparticle states are also protected by symmetry and topology.
Correspondingly, if we create exceptional points from an exceptional line by breaking the symmetry externally in the two-dimensional system,
the drumhead bulk states change into bulk Fermi arc states.

\section{Lattice models} \label{lattice}

\subsection{Exceptional surface in a $PT$-symmetric diamond lattice}
To obtain exceptional surfaces, we use a tight-binding model on a diamond lattice with $PT$ symmetry.
We write the three translation vectors 
as $\bm{t}_1=\frac{a}{2}(0,1,1)$, $\bm{t}_2=\frac{a}{2}(1,0,1)$, and $\bm{t}_3=\frac{a}{2}(1,1,0)$, where $a$ is the lattice constant.
The tight-binding model is
\begin{align}
	H=\sum _{<ij>}t_{ij}c_i^{^\dagger}c_j+i\sum _i\lambda _i c_i^{\dagger}c_i,
	\end{align}
where $t_{ij}$ and $\lambda _i$ are real constants.
The first term is nearest-neighbor hoppings between the sublattices A and B.
We assume that the hopping in direction of $\bm{\tau}=\frac{a}{4}(1,1,1)$ is different from the other three nearest-neighbor hoppings. 
We denote the nearest-neighbor hopping in the direction of $\bm{\tau}$ by $t_{\tau}$, and
the other hoppings by $t$.
The second term represents gain and loss leading to non-Hermiticity.
$\lambda _i$ takes values $+\lambda (-\lambda)$ for the A(B) sublattices. 
The model without the second term has been studied as a nodal-line semimetal \cite{Takahashi13, Okugawa17}.
The Hamiltonian in the momentum space is
\begin{align}
	H(\bm{k})=\Bigl(t_\tau +t\sum _i\cos (\bm{k}\cdot \bm{t}_i)\Bigr) \sigma _x
	+t\sum _i\sin (\bm{k}\cdot \bm{t}_i)\sigma _y+i\lambda \sigma _z.
\end{align}
Here, $\sigma _{x,y,z}$ are Pauli matrices acting on the sublattices.
In this model, the $PT$ operator is given by $PT=\sigma _xK$.

\begin{figure}[t]
	\includegraphics[width=8.5cm]{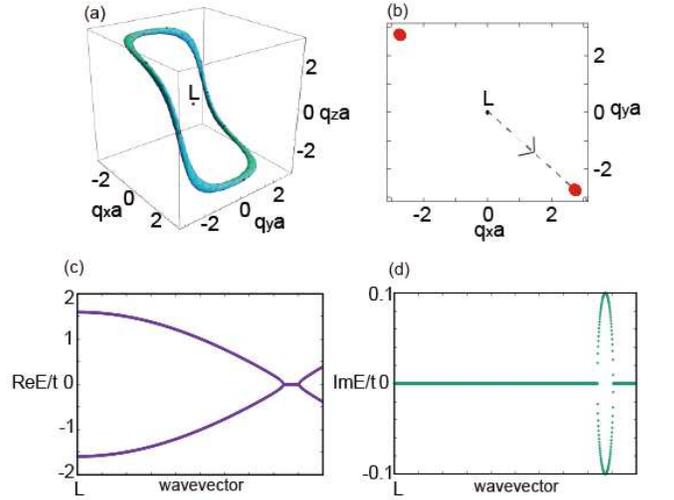}
	\caption{\label{diamond}(a) The exceptional surface around the $L$ point in the diamond lattice with $t_{\tau}=1.4t$ and $\lambda =0.1t$.
	$\bm{q}$ is a wavevector measured from the point $L$.	
	(b) The topological invariant $s(\bm{q})$ in the $q_z=0$ plane. The red and the white regions represent the areas with $s(\bm{q})=+1$ and $-1$, respectively.
    (c) and (d) The real and the imaginary parts of the energy bands along the dotted arrow in (b).}
\end{figure}

We set $t_{\tau}=1.4t$ and $\lambda =0.1t$ to investigate an exceptional surface.
The toruslike exceptional surface appears around the point $L=\frac{\pi}{a}(1,1,1)$,
as shown in Fig.~\ref{diamond}(a).
Then, the topological invariant $s(\bm{k})$ changes in the momentum space.
Figure \ref{diamond}(b) shows change in $s(\bm{k})$ by the exceptional surface.
The red regions with $s=+1$ give purely imaginary energy eigenvalues [Figs.~\ref{diamond}(c) and \ref{diamond}(d)].
These results agree with the discussions in Secs.~\ref{BD} and \ref{Z2}.

\subsection{Exceptional lines in a two-dimensional even-parity superconductor}
We study even-parity non-Hermitian superconductors with exceptional surfaces.
We consider a Bogoliubov-de Gennes Hamiltonian given by
$H=\frac{1}{2}\sum _{\bm{k}}\Psi ^{\dagger}_{\bm{k}}\mathcal{H}(\bm{k})\Psi _{\bm{k}}$
with $\Psi _{\bm{k}}^{\dagger}=(c_{\bm{k}\uparrow}^{\dagger},c_{\bm{k}\downarrow}^{\dagger},c_{-\bm{k}\uparrow},c_{-\bm{k}\downarrow})$
and 
\begin{align}
	\mathcal{H}(\bm{k})
	=\begin{pmatrix}
		\xi _{\bm{k}} & 0 & 0 & \Delta (\bm{k}) \\
		0 & \xi _{\bm{k}} & -\Delta (\bm{k}) & 0 \\
		0 & -\tilde{\Delta} (\bm{k}) & -\xi _{\bm{k}} & 0 \\
		\tilde{{\Delta}}(\bm{k}) & 0 & 0 & -\xi _{\bm{k}}
	\end{pmatrix}, \label{BdG}
\end{align}
where $\xi _{\bm{k}}$ is a kinetic energy satisfying $\xi _{\bm{k}}=\xi _{-\bm{k}}$,
and $\Delta (\bm{k})$ and $\tilde{\Delta}(\bm{k})$ are gap functions.
When $\tilde{\Delta}(\bm{k}) \neq \Delta ^{*}(\bm{k})$,
the superconductor becomes non-Hermitian.
The energy eigenvalues are given by $E_{\bm{k}}=\pm \sqrt{\xi _{\bm{k}}^2+\Delta(\bm{k})\tilde{\Delta}(\bm{k})}$.
In the even-parity superconductors,
$\Delta (\bm{k})=\Delta (-\bm{k})$ and $\tilde{\Delta}(\bm{k}) = \tilde{\Delta}(-\bm{k})$.
Such non-Hermitian superconductors can be realized if the pairing potential is complex \cite{Ghatak18}.

Because of $SU(2)$ symmetry,
the superconducting Hamiltonian in Eq.~(\ref{BdG}) can be rewritten as two $2\times 2$ matrices.
Here, we assume that $\tilde{\Delta}(\bm{k})=-\Delta ^{*}(\bm{k})$ to realize non-Hermiticity. 
Then, we obtain the block-diagonalized Hamiltonians as
\begin{align}
\mathcal{H}_{\pm}(\bm{k})
=\pm \mathrm{Im}[\Delta (\bm{k})]i\tau _x \pm \mathrm{Re}[\Delta (\bm{k})]i\tau _y +\xi _{\bm{k}}\tau _z,
\end{align}
where $\tau _{x,y,z}$ are Pauli matrices.
The energy eigenvalues are $E_{\bm{k}}=\pm \sqrt{\xi _{\bm{k}}^2-|\Delta (\bm{k})|^2}$,
which define exceptional surfaces as $\xi _{\bm{k}}^2=|\Delta (\bm{k})|^2$.
$\mathcal{H}_{\pm}$ have particle-hole symmetry $C=\tau _xK$ and inversion symmetry $P=1$.
Therefore, we can use the topological invariant $s'$
to characterize the exceptional surfaces for each of $\mathcal{H}_{\pm}$
although the bands of $\mathcal{H}$ are doubly degenerate.

\begin{figure}[t]
	\includegraphics[width=8.5cm]{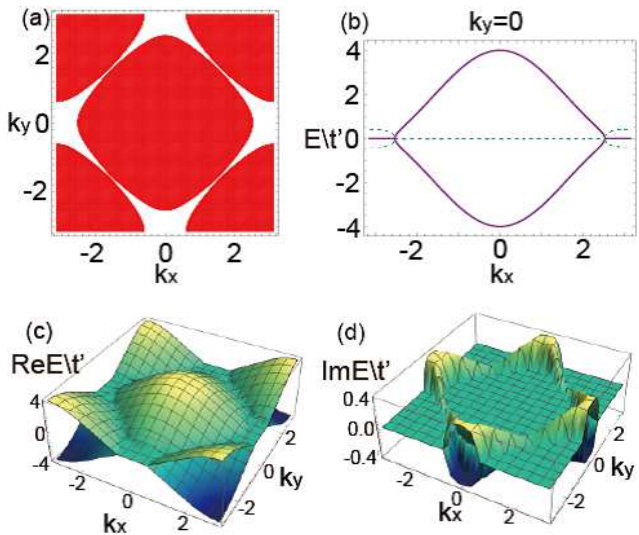}
	\caption{\label{SC}(a) The topological invariant $s'(\bm{k})$ for $\mathcal{H}_{+}$ in the momentum space.
		The red (white) regions indicate $s'(\bm{k})=+1 (-1)$.
		The boundaries between the two regions correspond to the exceptional lines.
	(b) The solid (dotted) lines are the real (imaginary) part of the energy bands on the $k_y=0$ line.
	(c) and (d) The real and the imaginary parts of the energy bands.}
\end{figure}

In this model, we put $\xi _{\bm{k}}=2t'(\cos k_x +\cos k_y)$ and
$\Delta(\bm{k})=\Delta _{x^2-y^2}(\cos k_x-\cos k_y)+i\Delta _{xy}\sin k_x\sin k_y$,
as an example of the two-dimensional even-parity superconductors.
The hopping $t'$ and the gap functions $\Delta _{x^2-y^2}$ and $\Delta _{xy}$ are real.
We set the lattice constant to unity.
Figure \ref{SC} shows exceptional lines in this model with $\Delta _{x^2-y^2}=0.2t'$ and $\Delta _{xy}=0.1t'$.
From Fig.~\ref{SC} (a), we find the exceptional lines as changes of $s'(\bm{k})$,
which are depicted by the boundaries between the red and the white regions in the momentum space.
The real and the imaginary parts of the energy become zero in the regions with $s'=-1$ and $s'=+1$ in Fig.~\ref{SC} (a), respectively.
As seen in Fig.~\ref{SC} (b)-\ref{SC}(d),
the non-Hermitian superconductor has topological nodes even if $\Delta (\bm{k})\neq 0$.

\section{Conclusion and Discussion}\label{conc}
In this paper, we have studied exceptional surfaces in non-Hermitian systems with $PT$ and $CP$ symmetries.
We have shown that the exceptional surfaces are topologically stable when $PT$ or $CP$ symmetry is present.
The results can be universally applied to $d$-dimensional non-Hermitian systems when $d\geq 1$.
The exceptional surfaces yield novel zero-gap quasiparticle states.
On the other hand, if the symmetry is broken by perturbations, the exceptional surfaces disappear.
Our theory is also confirmed by our calculation based on the lattice models.

From our study, we can see how to realize topological exceptional surfaces.
As a Dirac cone in two-dimensional systems becomes an exceptional line
by $PT$-symmetric non-Hermitian perturbations \cite{Szameit11, Ramezani12, Zhen15},
a two-dimensional exceptional surface can be obtained from a topological nodal line in three-dimensional systems.
Because a photonic crystal with a nodal line is suggested theoretically \cite{Lu13},
photonic systems have the potential for experimental realization of the exceptional surface.
Meanwhile, topological phase transitions typically need band touching in the bulk.
If the band touching happens by tuning parameters,
an exceptional surface can appear in the $PT$- and the $CP$-symmetric systems.
Therefore, when the system has the exceptional surface,
the phase may be regarded as a non-Hermitian intermediate phase
between trivial and topological gapped phases.
Moreover, 
because four-dimensional quantum Hall effects have been proposed by using synthetic dimension \cite{Price15, Price16, Ozawa16, Ozawa17},
non-Hermitian topological exceptional surfaces in higher-dimensional systems are feasible.

\textit{Note added}. Recently, we became aware of a related work \cite{Budich18}, 
which studies exceptional surfaces in view of Bernard and LeClair classes.
Additionally, $PT$-symmetric photonic crystals with exceptional surfaces are also theoretically proposed \cite{Zhou18a}.

\begin{acknowledgments}
This work was supported by World Premier International Reserach Center Initiative (WPI), MEXT, Japan, JPSJ Grant-in-Aid Scientific Research on Innovative Areas "Discrete Geometric Analysis for Materials Design" (Grant No. 17H06460), and "Nano Spin Conversion Science" (Grant No. JP17H05179), and the JSPS-EPSRC Core-to-Core program "Oxide Superspin".
\end{acknowledgments}

\bibliographystyle{apsrev4-1}
\bibliography{NH}

\begin{thebibliography}{97}
\bibitem [{com()}]{comment}%
  \BibitemOpen
  \href@noop {} {\bibinfo  {journal} {Because the $\mathbb{Z}_2$ topological
  invariant classifies the zero-dimensional systems, the characterization is
  applicable to general $(d-1)$-dimensional Fermi surfaces as well as the
  exceptional surfaces in the non-Hermitian $PT$-symmetric systems. We can
  focus on the exceptional surface by ignoring the term of the identity matrix
  in the effective model}\ }\BibitemShut {NoStop}%
\end{thebibliography}

\end{document}